\documentclass[prb, showpacs]{revtex4}
\usepackage{amssymb}
\usepackage{bm}
\usepackage{graphicx}%
\begin{document}
\title
{Charge ordering and interlayer phase coherence in quantum Hall
superlattices}
\author{S.\,I.\,Shevchenko}
\affiliation{%
B.\,I.\, Verkin Institute for Low Temperature Physics and
Engineering National Academy of Sciences of Ukraine, Lenin av. 47
Kharkov 61103, Ukraine}
\author{D.\,V.\,Fil}
\affiliation{ Institute for Single Crystals, National Academy of
Sciences of Ukraine, Lenin av. 60, Kharkov 61001, Ukraine}
\author{A.\,A.\,Yakovleva}
\affiliation{%
National Technical University "Kharkov Polytechnical Institute,"
ul. Frunze 21, Kharkov 61002, Ukraine}


\begin{abstract}
The possibility of the existence of states with a spontaneous
interlayer phase coherence in multilayer electron systems in a
high perpendicular to the layers magnetic field is investigated.
It is shown that phase coherence can be established in such
systems only within individual pairs of adjacent layers, while
such coherence does not exist between layers of different pairs.
The conditions for stability of the state with interlayer phase
coherence against transition to a charge-ordered state are
determined. It is shown that in the system with the number of
layers $N\leq 10$ these conditions are satisfied  at any value of
the interlayer distance $d$. For $N>10$ there are two intervals of
stability: at sufficiently large and at sufficiently small $d$.
For $N\to \infty$ the stability interval in the region of small
$d$ vanishes.
\end{abstract}


\maketitle

\section{Introduction}
In the last 10 years main attention in studying  the quantum Hall
effect and related phenomena was given to bilayer systems. One of
the most interesting features of bilayer quantum Hall systems is
the possibility of realizing in them a superfluid state of a gas
of electron-hole pairs with spatially separated components. For
the first time, the possibility of superfluidity of this type  was
predicted in Refs. \onlinecite{1} and \onlinecite{2} with
reference to bilayer electron-hole systems in zero magnetic field
(see also Refs. \onlinecite{3} and \onlinecite{4}). In Ref.
\onlinecite{5} (see also Refs. \onlinecite{6,7,8,9} for more
details) it was shown that strong magnetic field applied to a
bilayer electron-hole system favors the formation of a superfluid
state, since, unlike the situation considered in Refs.
\onlinecite{1} and \onlinecite{2}, in the latter case it is not
necessary to satisfy the condition of nesting of the Fermi
surfaces of the electrons and holes.

Owing to the electron-hole symmetry, a bilayer electron-electron
system in a quantizing magnetic field with filling factors of the
layers equal to $\nu_1=\nu$ and $\nu_2=1-\nu$ ($\nu<1$) is a
practically complete analog of  bilayer electron-hole systems, and
a transition to a superfluid phase is possible in such a system as
well\cite{10,11,12,13,14,15,16}. Recently, interest in this
question has risen significantly in connection with
experiments\cite{17,18} in which an effect of the Josephson type
was observed (a sharp peak of the differential tunneling
conductance at  gate voltage biased near zero). Furthermore, in a
recent experiment\cite{19} on the interlayer drag in bilayer
systems a strong suppression of the longitudinal drag current and
an appearance of the Hall component of the drag current was
observed. This effect takes place in the situation when the total
filling factor $\nu_{tot}=\nu_1+\nu_2$ becomes close to unity. The
effects observed in Refs. \onlinecite{17,18,19} can be explained
on the assumption that a spontaneous interlayer phase coherence
arises in the system and a transition of the system to a
superfluid state occurs. A direct experimental proof  of
Bose-Einstein condensation of interwell excitons (photoexcited
electron-hole pairs with spatially separated components) in double
quantum wells in zero magnetic field was also obtained recently
\cite{20}. The proof\cite{20} was based on the measured
temperature dependence of the luminescence spectra.

For further understanding of this phenomenon it is naturally to
study the superfluid properties of multilayer electron systems.
Multilayer systems are of particular interest for studying
collective properties of electron-hole superfluids by optical and
acoustic methods, since the integrated intensity of the
interaction of such systems with external fields increases in
proportion to the number of layers. The present study is devoted
to investigation of the effect of interlayer phase coherence in
quantum Hall superlattices.

The realization of multilayer quantum Hall systems was reported in
Refs. \onlinecite{21,22,23}. The research\cite{21,22,23} was
mainly concerned with analysis of the properties of chiral edge
states in such systems. Previously, the question of interlayer
phase coherence in multilayer quantum Hall systems was considered
in Refs. \onlinecite{24} and \onlinecite{25}. The authors of Ref.
\onlinecite{25} have shown that a state in which phase coherence
arises between all layers can be realized in such systems (below,
this is referred to  as a state with global phase coherence). It
has been shown in Ref. \onlinecite{24} that in a multilayer
quantum Hall system another state can arise (which we shall call a
dimer state) in which the system separates into pairs of adjacent
layers and the interlayer phase coherence is established only
within each pair, while there is no such coherence  between layers
belonging to different pairs. This raises the question of
reconciliation of the results of Refs. \onlinecite{24} and
\onlinecite{25}. In the present paper we show that the states of
Refs. \onlinecite{24} and \onlinecite{25} correspond to two
different solutions of the self-consistency equations for the
order parameters describing the interlayer phase coherence. We
find that the energy of the dimer state is less than the energy of
the state found in Ref. \onlinecite{25} and, hence, it is the
dimer state that should be considered as a candidate for the
ground state of such a system.

It was shown in Ref. \onlinecite{25} that  conditions for the
existence of the state with interlayer phase coherence in
multilayer systems are more restrictive than for bilayer systems.
Multilayer quantum Hall systems, unlike the bilayer ones,
demonstrate a tendency toward a formation of a charge-ordered
state (where the filling factors of adjacent layers become equal
to 0 and 1). In Ref. \onlinecite{25} the stability conditions for
the state with interlayer phase coherence were analyzed under the
assumption that the phase coherence has a global character. Since
this assumption is not confirmed, the stability conditions will
have to be reexamined. We find that in the limit of an infinite
number of layers (the case considered in Ref. \onlinecite{25}) the
dimer state is stable against transition to the charge-ordered
state already at interlayer distances $d>d_c\approx 1.45\ell_B$
($\ell_B$ is the magnetic length), which is less than the critical
value ($d_c\approx 1.7\ell_B$ ) obtained in Ref. \onlinecite{25}.

In connection with the fact that the properties of systems with
number of layers $N= 2$ and $N\to \infty$ are substantially
different, and in real physical systems the number of layers is
always finite, an important question arises as to how the
stability conditions depend on $N$. In this paper this question is
studied with reference to the system with even $N$. It is found
that for $N\leq 10$ the state with interlayer phase coherence is
stable for arbitrary $d$. For $N
>10$ stability takes place only for $d<d_{c1}$ or for $d>d_{c2}$, where
$d_{c1}$ decreases and $d_{c2}$ increases with increasing $N$
while remaining in the interval $0<d_{c1} <d_{c2}<d_c$.

\section{Interlayer phase coherence in  bilayer system with total
filling factor $\nu_{tot} =1$}

Prior studying the multilayer system, we describe the approach
used in this paper with reference to the simplest case $N=2$ and
recall the bilayer situation. Let us consider a double-layer
electron system subjected by a high magnetic field applied
perpendicular to the layers. We imply the filling factor
$\nu_{tot} = \nu_1 + \nu_2=1$. Tunneling between layers is assumed
to be so weak to be neglected in the Hamiltonian. If the cyclotron
frequency is much higher than all the other characteristic
energies of the problem, the system  can be treated in the lowest
Landau level approximation, in which the influence of the upper
unfilled Landau levels on the dynamics of the system is neglected.

The Hamiltonian of this system has the form
\begin{eqnarray}\label{1}
  H=\frac{1}{2}\sum_{n,n'=1,2} \int \int d^2 r_n d^2 r_{n'}
  \Psi^+_n({\bf r}_n)\Psi^+_{n'}({\bf r}_{n'}) V_{nn'}(|{\bf r}_n-
  {\bf r}_{n'}|)\Psi_{n'}({\bf r}_{n'})\Psi_n({\bf r}_n)\cr
   -
  \sum_{n=1,2} \int d^2 r_n \mu_n \Psi^+_n({\bf r}_n) \Psi_n({\bf
  r}_n) +H_{BG},
\end{eqnarray}
where  $\Psi_n$ are the Fermi field operators,
\begin{equation}\label{2}
  V_{nn'}(r)=\frac{e^2}{\varepsilon\left[r^2+(n-n')^2
d^2\right]^{1/2}} \ -
\end{equation}
is the Coulomb potential, $\varepsilon$,  the dielectric constant,
and $\mu_n$, the chemical potential in layer $n$. The term
$H_{BG}$ in (\ref{1}) describes the interaction of electrons with
positively charged impurities in doping layers. For simplicity we
assume that the z coordinates of the electron layers and doping
layers coincide. If one takes into account the difference of the z
coordinates of the electron and doping layers which takes place in
the real physical situation it results in an appearance of
additional constant term in the energy that does not influence the
effects studied in this paper.

For further analysis it is convenient to transform to the
operators of the Fourier components of the electron density
$\rho_n({\bf q})$, in terms of which the Hamiltonian (\ref{1}) can
be written in the form
\begin{eqnarray}\label{3}
  H=\frac{1}{2 S}\sum_{n,n'=1,2} \sum_{\bf q} \left\{
 V_{nn'}(q)\left[ \rho_n({\bf q})\rho_{n'}(-{\bf
 q})-\delta_{nn'}\exp(-\frac{q^2\ell_B^2}{2})\rho_n(0)\right]\right\}\cr-
 \sum_{n=1,2} \mu_n \rho_n (0)+H_{BG},
\end{eqnarray}
where $S$ is the area of the layer. The operator  $\rho_n({\bf
q})$ expressed in terms of the creation and annihilation operators
for electrons in the lowest Landau level ($a^+(X)$, $a(X)$) ($X$
is the guiding center of the orbit) has the form:
\begin{equation}\label{4}
  \rho_n({\bf q})=\sum_X a_n^+(X+q_y\ell_B^2/2)a_n(X-q_y\ell_B^2/2)\exp(iq_x X-
  q^2\ell_B^2/4),
\end{equation}
In Eq. (\ref{3}) the quantity
\begin{equation}\label{5}
  V_{nn'}(q)=\frac{2\pi e^2}{\varepsilon q} \exp(-q d |n-n'|)
\end{equation}
is the Fourier component of the Coulomb potential.

For further analysis we use the mean field approximation. In this
approximation the Hamiltonian (\ref{3}) takes the form
\begin{equation}\label{6}
  H_{MF}= \sum_X \left\{\epsilon_1 a^+_1(X) a_1(X)+
  \epsilon_2 a^+_2(X) a_2(X) -[J\Delta a^+_1(X) a_2(X)+ h.c.]\right\}.
\end{equation}
The energies $\epsilon_n$ in (\ref{6}) are
\begin{equation}\label{7}
  \epsilon_n=V_n-I\nu_n-\mu_n ,
\end{equation}
where $V_n$ are the Fourier components of the Coulomb interaction
at q = 0  screened by positively charged doping impurities. For
the bilayer system these quantities are equal to
\begin{eqnarray}\label{8}
  &&V_1=\frac{1}{2\pi\ell_B^2}\lim_{q\to
  0}\left[V_{11}(q)\left(\nu_1-\frac{1}{2}\right)+
  V_{12}(q)\left(\nu_2- \frac{1}{2}\right)\right]=W\tilde{\nu},
  \cr
  &&V_2=\frac{1}{2\pi\ell_B^2}\lim_{q\to
  0}\left[V_{22}(q)\left(\nu_2-\frac{1}{2}\right)+
  V_{21}(q)\left(\nu_1- \frac{1}{2}\right)\right]=-W \tilde{\nu}. \
  \end{eqnarray}
The parameter $\tilde{\nu}=(\nu_1-\nu_2)/2$ describes the value of
the imbalance of the filling factors of the layers. In Eq.
(\ref{6})  $\Delta$ is the order parameter, defined as
$\Delta=\langle a_2^+(X) a_1(X)\rangle$. In general case the order
parameter $\Delta=|\Delta|e^{i\varphi}$ is complex valued. We
restrict consideration to the case where its modulus $|\Delta|$
and phase $\varphi$  are independent of X. In deriving Eq.
(\ref{6}) we have also taken into account that the averages
$\langle a_n^+(X) a_n(X)\rangle$ are equal to $\nu_n $. In Eqs.
(\ref{6}), (\ref{7}), and (\ref{8}) we have used the following
energy parameters: the parameter
\begin{equation}\label{9}
  W=\frac{e^2 d}{\varepsilon\ell_B^2}
\end{equation}
that describes the energy of the direct Coulomb interaction
between layers, and the parameters
\begin{equation}\label{10}
  I=\frac{1}{S} \sum_{\bf q} V_{11}(q)
  \exp(-q^2\ell_B^2/2)=\sqrt{\frac{\pi}{2}}\frac{e^2}{\varepsilon\ell_B},
\end{equation}
\begin{equation}\label{11}
J=\frac{1}{S} \sum_{\bf q} V_{12}(q)
\exp(-q^2\ell_B^2/2)=\sqrt{\frac{\pi}{2}}\frac{e^2}{\varepsilon\ell_B}\exp
\left(\frac{d^2}{2 \ell_B^2}\right){\rm
erfc}\left(\frac{d}{\ell_B\sqrt{2}}\right)
\end{equation}
that describe the energy of the intralayer and interlayer exchange
interactions.

In a state in which the order parameter $\Delta$ is nonzero, the
motion of electrons in one layer is correlated with the motion of
holes (empty single-particle states on the lowest Landau level) in
the other layer. This state can be treated as a gas of
electron-hole pairs, which are composite bosons. The Bose-Einstein
condensate of such pairs (a true condensate at $T = 0$ or a
quasi-condensate with a fluctuating phase at $T\ne 0$) will
demonstrate superfluid properties. The state with nonzero
superflow is the state in which nondissipative electrical current
in one layer is accompanied by equal and oppositely directed
nondissipative electrical current in the other layer.

For finding the self-consistency conditions for the order
parameter $\Delta$ and chemical potentials $\mu_1$ and $\mu_2$ we
subject the operators $a_1$ and $a_2$ to a $u-v$ transformation of
the form
\begin{eqnarray}\label{12}
  a_1=u\alpha+v^*\beta^+ ,\cr
  a_2=u^*\beta^+ - v\alpha .
\end{eqnarray}
The operators $\alpha$ and  $\beta$ will satisfy the Fermi
commutation relations if  $|u|^2+|v|^2=1$. This allows  to seek
the coefficients of the $u-v$ transformation in the form
$u=\cos(\Theta/2)$ and $v=\sin(\Theta/2)e^{i\chi}$. From the
requirement of vanishing the terms nondiagonal in $\alpha$ and
$\beta$ in the transformed Hamiltonian, we find that
\begin{equation}\label{13}
  \sin\Theta=\frac{J|\Delta|}{\sqrt{\tilde{\epsilon}^2+(J|\Delta|)^2}},\qquad
 \cos\Theta=\frac{\tilde{\epsilon}}{\sqrt{\tilde{\epsilon}^2+(J|\Delta|)^2}},
\end{equation}
and that $\chi$ coincides with the phase $\varphi$ of the order
parameter. Here we introduce the notation
$\tilde{\epsilon}=(\epsilon_1-\epsilon_2)/2$.

Using Eqs. (\ref{12}) and (\ref{13}), we obtain the following
self-consistency equations:
\begin{eqnarray}\label{14}
  \nu_1+\nu_2=1=\langle a^+_1 a_1\rangle+\langle a^+_2
  a_2\rangle=1+N_F({\cal E}_\alpha)-N_F({\cal E}_\beta),
\end{eqnarray}
\begin{eqnarray}\label{15}
  2\tilde{\nu}=\langle a^+_1 a_1\rangle-\langle a^+_2
  a_2\rangle=-\frac{\tilde{\epsilon}}
  {\sqrt{\tilde{\epsilon}^2+(J|\Delta|)^2}}[1-N_F({\cal E}_\alpha)-N_F({\cal
  E}_\beta)],
\end{eqnarray}
\begin{eqnarray}\label{16}
  \Delta=\langle a^+_2 a_1\rangle=\frac{J\Delta}
  {2\sqrt{\tilde{\epsilon}^2+(J|\Delta|)^2}}[1-N_F({\cal E}_\alpha)-N_F({\cal
  E}_\beta)],
\end{eqnarray}
where
\begin{equation}\label{17}
  {\cal E}_{\alpha(\beta)}=\sqrt{\tilde{\epsilon}^2+
  (J|\Delta|)^2}\pm\frac{\epsilon_1+\epsilon_2}{2} \ -
\end{equation}
are the energies of elementary excitations and  $N_F({\cal
E})=[\exp({\cal E}/T)+1]^{-1}$ is a the Fermi distribution
function. It follows from Eq. (\ref{14}) that ${\cal
E}_\alpha={\cal E}_\beta$ , which leads to the condition
$\epsilon_1+\epsilon_2=0$. From Eqs. (\ref{15}) and (\ref{16}) one
can easily find the dependence of the modulus of the order
parameter on the value of the imbalance and temperature. Below we
specify the case of zero temperatures, when this dependence has
the form
 \begin{equation}\label{18}
  |\Delta(\tilde{\nu})|=\sqrt{\frac{1}{4}-\tilde{\nu}^2}.
\end{equation}

The differential of the free energy of the system at
$\nu_1+\nu_2=1$ and $T = 0$ is equal to
\begin{equation}\label{19}
  d F= \mu_{1} d \nu_1+\mu_{2} d\nu_2=
  2\tilde{\mu}d\tilde{\nu},
\end{equation}
where  $\tilde{\mu}=(\mu_1-\mu_2)/2$. Here and below we give the
energy per electron. If the analytical form of the function
$\tilde{\mu}(\tilde{\nu})$ is known, the energy of the system as a
function of the imbalance value is easily found:
\begin{equation}\label{20}
  F(\tilde{\nu})=F(-1/2)+2\int_{-1/2}^{\tilde{\nu}} d\tilde{\nu}'
  \tilde{\mu}(\tilde{\nu}'),
\end{equation}
where $F(-1/2)$ is the energy of the system for
$\tilde{\nu}=-1/2$, i.e., for  $\nu_1=0$ and $\nu_2=1$. At such
filling factors the energy of the system is the sum of the
energies of the direct Coulomb interaction between layers and the
intralayer exchange interaction in layer 2, i.e.,
$F(-1/2)=W/4-I/2$. Using Eq. (\ref{7}) and taking into account the
relation $\tilde{\epsilon}=-J\tilde{\nu}$ that follows from Eqs.
(\ref{15}) and (\ref{16}), we find the following expression for
$\tilde{\mu}$: $$\tilde{\mu}( \tilde{\nu})=(W-I+J)\tilde{\nu}.$$
Accordingly,
\begin{equation}\label{21}
  F(\tilde{\nu})=
   -\frac{I+J}{4}+(W-I+J)\tilde{\nu}^2.
\end{equation}
It is  useful to present expression (\ref{21}) in the form
\begin{equation}\label{22}
  F=
    W\tilde{\nu}^2-\frac{I}{2}[\nu_1^2+(1-\nu_1)^2]-J|\Delta|^2
\end{equation}
from which the physical meaning of each of the terms becomes
obvious. The first term is the energy of the direct Coulomb
interaction at  filling factor imbalance $\tilde{\nu}$. The second
term is the sum of the energies of the intralayer exchange
interaction in layers 1 and 2. The third term is the energy
connected with the interlayer phase coherence.

Since energy (\ref{22}) is independent of the phase $\varphi$ of
the order parameter, the interlayer phase coherence is spontaneous
one. The minimum energy of a state with interlayer phase coherence
is reached for the modulus and phase of the order parameter
independent of the coordinate. A state with nonzero superfluid
current that corresponds to nonzero gradients of the phase is
higher in energy, but in the case where the gradient of the phase
is less than certain critical value (which depends on the ratio
$d/\ell_B$), such a state remains  metastable, because the system
cannot emit normal excitations, i.e., the Landau criterion of
superfluidity is satisfied.

To conclude this Section we turn our attention to an important
observation that follows from formula (\ref{21}). Using Eqs.
(\ref{9})-(\ref{11}), we  see that the coefficient of the
$\tilde{\nu}^2$  term on the right-hand side of (\ref{21}) is
positive for arbitrary $d\ne 0$. This plays a fundamental role for
the existence of the state with interlayer phase coherence in
bilayer systems. Indeed, if the sign of the expression $W-I+J$
were negative, then the system would go to a state with a maximum
imbalance of the filling factors, $\tilde{\nu}=\pm 1/2$, for which
$|\Delta|= 0$ (see Eq. (\ref{18})) and, consequently, interlayer
phase coherence is absent. Skipping  ahead, we note that in
multilayer systems, in contrast, such a mechanism of destroying
the phase coherence can indeed be realized in a certain range of
$d$. This question is analyzed in details in Sec. \ref{sec4}.

\section{STATES WITH INTERLAYER PHASE COHERENCE IN  MULTILAYER SYSTEM}

Let us consider a system with the number of layers $N\to \infty$
with an average filling factor per layer equal 1/2. The
Hamiltonian of such a system can be written in the form (\ref{3})
with the sum over two layers replaced by a sum over an infinite
number of layers. In the mean field approximation, the Hamiltonian
of the multilayer systems is reduced to
\begin{equation}\label{23}
  H_{MF}=\sum_n \sum_X\{\epsilon_n a_n^+ a_n - \frac{1}{2} \sum_{m\ne 0} [J_m \Delta(n,m) a^+_n
  a_{n+m}+h.c.]\},
\end{equation}
where the energy parameters $J_m$ are equal to
\begin{equation}\label{24}
  J_m=\sqrt{\frac{\pi}{2}}\frac{e^2}{\varepsilon\ell_B}\exp
\left(\frac{d^2 m^2}{2 \ell_B^2}\right){\rm erfc}\left(\frac{d |m|
}{\ell_B\sqrt{2}}\right)
\end{equation}
(the parameter $J$ introduced in the previous Section coincides
with $J_{±1}$) and the order parameters are defined as
\begin{equation}\label{25}
 \Delta(n,m) =\langle  a^+_{n+m}
  a_n\rangle
\end{equation}
In general case the quantities $\Delta(n,m)$ can be nonzero for
arbitrary $m$ (not only for $m = \pm 1$). This corresponds to the
situation when a correlation arises between the electrons
belonging not only to the adjacent layers but also to the layers
arbitrarily far apart. Since the Coulomb potential is long-ranged,
such correlations may give significant contribution to the energy,
and it is important to take them into account for finding the
ground state of the system.

It is natural to expect from physical reasons that if an imbalance
of the filling factors arise in such a system it has a periodic
character. Let us consider the situation when such a period does
not exceed twice the distance between adjacent layers. As it shown
below, the doubling of the period may happen even in the absence
of imbalance of the filling factors. Latter possibility
corresponds to the state with interlayer phase coherence of the
dimer type.

Let the filling factors of the layers be equal to
$\nu_n=1/2+(-1)^n\tilde{\nu}$. In this case the quantity
$\tilde{\nu}$ can be treated as a charge-ordering parameter.
Basing on the assumption of periodicity of the system, we seek the
order parameter $\Delta(n,m)$ in the form
\begin{equation}\label{26}
  \Delta(n,m)=\Delta_1(m)+(-1)^n
  \Delta_2(m).
\end{equation}

In general case the quantities $\Delta_{1(2)}(m)$ can be complex
valued, i.e., they can contain phase factors
$e^{i\varphi_{1(2)}(m)}$. The phase factors can no longer be
specified independently, since the self-consistency equation
imposes certain conditions on the phase difference
$\varphi_{1(2)}(m)$. However, these conditions do not fix all of
the phases, and a certain arbitrariness in the choice of the
quantities $\varphi_{1(2)}(m)$ remains. It can be shown that the
energy of the system is independent of the phases of the order
parameters if the latter satisfy the self-consistency equations
(we stress that we are considering the case when the modulus and
phase of the order parameter are independent of $X$). An
arbitrariness in the choice of phases is just a reflection of the
spontaneous character of the phase coherence. It allows us to be
more specific and restrict consideration to the simplest case of
all the phases are equal to zero and the quantities $\Delta_1(m)$
and $\Delta_2(m)$ are real valued.

Transforming in Eq. (23) to the Fourier components of the
operators  $a_n$
\begin{equation}\label{27}
  a_n=\frac{1}{\sqrt{N}}\sum_{q_z} e^{i q_z n} a_{q_z},
\end{equation}
we obtain the following expression for the Hamiltonian of the
system:
\begin{equation}\label{28}
  H_{MF}=\sum_X \sum_{q_z}\{[\epsilon - \Delta_1(q_z)] a_{q_z}^+ a_{q_z}
   + [\tilde{\epsilon}-\Delta_{2r}(q_z)
  -i \Delta_{2i}(q_z)] a^+_{q_z+\pi}a_{q_z}\}.
  \end{equation}
Here the following notations are introduced
$\epsilon=(\epsilon_{2k}+\epsilon_{2k-1})/2$,
$\tilde{\epsilon}=(\epsilon_{2k}-\epsilon_{2k-1})/2$. In view of
the assumed periodicity in z  the quantities $\epsilon$ and
$\tilde{\epsilon}$ are independent of k. The functions $
\Delta_1(q_z)$ and $\Delta_{2r(i)}(q_z)$ in (\ref{28}) are defined
as
\begin{equation}\label{29}
  \Delta_1(q_z)=2\sum_{m=1}^\infty J_m\Delta_1(m)\cos (m q_z),
\end{equation}
\begin{equation}\label{30}
  \Delta_{2r}(q_z)=2\sum_{m=1}^\infty J_{2 m}\Delta_2(2 m)\cos (2 m
  q_z),
\end{equation}
\begin{equation}\label{31}
  \Delta_{2i}(q_z)=2\sum_{m=0}^\infty J_{2m+1}\Delta_2(2m+1)\sin [(2m+1)
  q_z].
\end{equation}

Diagonalizing the Hamiltonian (\ref{28}) and calculating the
averages of (\ref{26}), we arrive to the self-consistency
conditions for the order parameters. Here we present these
equations for the case $T = 0$:
\begin{equation}\label{32}
  \tilde{\nu}=-\frac{1}{2\pi}\int_{-\pi/2}^{\pi/2} d q_z
  \frac{\tilde{\epsilon}-\Delta_{2r}(q_z)}{E(q_z)},
\end{equation}
\begin{equation}\label{33}
  \Delta_1(2m)=0,
\end{equation}
\begin{equation}\label{34}
  \Delta_1(2m+1)=\frac{1}{2\pi}\int_{-\pi/2}^{\pi/2} d q_z
  \frac{\cos[(2m+1) q_z] \Delta_1(q_z)}{E(q_z)},
\end{equation}
\begin{equation}\label{35}
  \Delta_2(2m)=-\frac{1}{2\pi}\int_{-\pi/2}^{\pi/2} d q_z
  \frac{\cos(2mq_z)[\tilde{\epsilon}-\Delta_{2r}(q_z)]}{E(q_z)},
\end{equation}
\begin{equation}\label{36}
  \Delta_2(2m+1)=\frac{1}{2\pi}\int_{-\pi/2}^{\pi/2} d q_z
  \frac{\sin[(2m+1)q_z]\Delta_{2i}(q_z)}{E(q_z)}.
\end{equation}
In equations (\ref{32}) and (\ref{34})-(\ref{36}) the spectrum of
energies of elementary excitations $E(q_z)$ is determined by the
following relation:
\begin{eqnarray}\label{37}
  E(q_z)=\sqrt{ [\tilde{\varepsilon}-\Delta_{2r}(q_z)]^2+
  \Delta_1^2(q_z)+\Delta_{2i}^2(q_z)}.
\end{eqnarray}

The state with global phase coherence  considered in Ref.
\onlinecite{25} corresponds to the solution of
(\ref{32})-(\ref{36}) with all the quantities $\Delta_2(2m+1)$
equal zero (in this case Eq. (\ref{36}) is satisfied
automatically) and $\Delta_1(2m+1)\ne 0$ for arbitrarily large
$m$. The latter  means that phase coherence is established between
all layers.

It is easy to see that the self-consistency equations
(\ref{32})-(\ref{36}) are also satisfied by a solution of an
essentially different type: $\Delta_1(\pm 1)=\pm\Delta_2(\pm 1)\ne
0$, but $\Delta_1(m)=\Delta_2(m)=0$ at  $m\ne \pm 1$. In this case
$\Delta(2k,1)=\Delta(2k+1,-1)\ne 0$ (or
$\Delta(2k-1,1)=\Delta(2k,-1)\ne 0$), but all the remaining
$\Delta(n,m)$ are zero. In the state corresponding to such a
solution, the system separates into pairs of layers, and the
interlayer phase coherence is established only within each pair.
We call this state the dimer state.

Writing the differential of the free energy of a multilayer system
and integrating it over $\tilde{\nu}$, we obtain an expression for
the free energy of the multilayer system:
\begin{equation}\label{38}
  F(\tilde{\nu})=
   V_c(\tilde{\nu})-I\left(\frac{1}{4}+\tilde{\nu}^2\right) -2\sum_{m=
   1}^\infty
   J_m[\Delta_1^2(m)+\Delta_2^2(m)].
\end{equation}
where the first term $V_c(\tilde{\nu})$ corresponds to the energy
of the direct Coulomb interaction, which is independent of the
concrete form of $\Delta_{1(2)}(m)$.

Using the answer (\ref{38}), let us compare the energies of the
state with global phase coherence and of the dimer state for
$\tilde{\nu}=0$. At zero imbalance Eqs. (\ref{32})-(\ref{36}) give
the following expressions for the order parameters
$\Delta_{1,2}(m)$. For a state with global coherence we have
\begin{equation}\label{39}
  \Delta_1(m)=\frac{\sin(\pi m/2)}{m\pi},\qquad
  \Delta_2(m)=0.
\end{equation}
For the dimer state $\Delta_1(m)=\pm\Delta_2(m)=\delta_{1,|m|}/4$.
Consequently, the energy difference of the dimer state and the
state with global phase coherence satisfies the inequality
\begin{equation}\label{40}
  F_d-F_g=-\frac{J_1}{4}+\frac{2}{\pi^2}\sum_{k=0}^\infty
  J_{2k+1}\frac{1}{(2k+1)^2}\leq -\frac{J_1}{4}\left(1-\frac{8}{\pi^2}\sum_{k=0}^\infty
  \frac{1}{(2k+1)^2}\right)=0.
\end{equation}
In deriving Eq. (\ref{40}) we take into account that for $d\ne 0$
the parameters $J_m$ the smaller the large $|m|$. For $d= 0$, when
all  $J_m$ are identical, the energies of the two states under
consideration are equal. A numerical analysis of the
self-consistency equations shows that the inequality $F_d<F_g$
holds for arbitrary values of $\tilde{\nu}$ and $d\ne 0$. Since
the energy of the state with global phase coherence is higher than
the energy of the dimer state,  this state but not the state
considered in Ref. \onlinecite{25}  should be considered as a
possible candidate for the ground state of the system.

\section{COMPARISON OF THE ENERGIES OF THE DIMER AND
CHARGE-ORDERED STATES} \label{sec4}

Let us now consider the question of the stability of the dimer
state in a multilayer system with even number of layers against a
transition to the charge-ordered state. We should specify the
explicit form of the energy of the direct Coulomb interaction
$V_c(\tilde{\nu})$. It can be easily done by using the fact that
this energy is equal to the energy of the electric field induced
inside the superlattice due to deviation of the filling factors of
individual layers from the average value. Since the system on the
whole remains electrically neutral, the electric field outside the
superlattice is equal to zero.

For instance, for the filling factors of the layers equal to
$\nu_{2k-1}=1/2-\tilde{\nu}$ and $\nu_{2k}=1/2+\tilde{\nu}$
($k=1,2,\ldots,N/2$), the electric field is nonzero between layers
1 and 2, 3 and 4, 5 and 6, etc., while between layers 2 and 3, 4
and 5, etc. it equals to zero. Then, the electric field energy per
electron, as in the bilayer system, would be equal to
$V_c=W\tilde{\nu}^2$. However, in contrast to the bilayer system,
in the multilayer system substantial benefit in the energy of the
direct Coulomb interaction can be achieved by changing the filling
factors of only the two outer layers. Indeed, if the filling
factors of the outer layers take the values
$\nu_{1}=1/2-\tilde{\nu}/2$ and $\nu_{N}=1/2+\tilde{\nu}/2$, then
the electric field is induced between all the layers, and its
absolute value would be smaller by a factor of two than in the
previous case (the direction of the field would change sign on
crossing a layer). Consequently, in the last case the energy $V_c$
(energy per electron) is in two times smaller than in the first
case (with an accuracy up to $1/N$). In such a state the total
energy is
\begin{equation}\label{41}
  F=-\frac{I+J}{4}+\left(\frac{W}{2}-I+J\right)\tilde{\nu}^2.
\end{equation}
In deriving Eq. (\ref{41}) we take into account that in the dimer
state the order parameter $\Delta=2\Delta_{1(2)}$ has the same
dependence on $\tilde{\nu}$ as in the bilayer system (Eq.
(\ref{18})). In contrast to Eq. (\ref{21}), the sign of the
coefficient of the $\tilde{\nu}^2$ term in Eq. (\ref{41}) depends
on $d$: for $d < 1.45\ell_B$ this coefficient is negative, while
at larger $d$ it is positive. Consequently, for $d < 1.45\ell_B$
and $N\to \infty$ the minimum energy  corresponds to the state
with the maximum charge ordering, in which all the  $\Delta(n,m)$
go to zero and phase coherence is absent. In the opposite case $d
> 1.45\ell_B$ the minimum of energy (\ref{41}) is reached for $\tilde{\nu}=0$
and charge ordering becomes energetically unfavorable.

The above analysis is easily generalized to the case of finite
even $N$. Let us consider the configuration where the inner layers
are in a completely or partially charge-ordered state, and for
these layers the imbalance of the filling factors is equal to
$\tilde{\nu}_{in}$. The imbalance of the filling factors of the
outer layers $\tilde{\nu}_{ex}=({\nu}_N-{\nu}_{1})/2$ can be
different from $\tilde{\nu}_{in}$. We imply that phase coherence
is established only between the inner layers and has the dimer
character. The energy (per electron) of such a state is equal to
\begin{eqnarray}\label{42}
  F(\tilde{\nu}_{ex},\tilde{\nu}_{in})= W
  \left[\tilde{\nu}_{ex}^2 +\frac{N-2}{N}
  \left(\tilde{\nu}_{in}-\tilde{\nu}_{ex}\right)^2\right]- I
    \left[\frac{N-2}{N}\left(\frac{1}{4}+\tilde{\nu}_{in}^2\right)
  +\frac{2}{N}\left(\frac{1}{4}+\tilde{\nu}_{ex}^2\right)\right] -J
  \frac{N-2}{N}\left(\frac{1}{4}-\tilde{\nu}_{in}^2\right).
\end{eqnarray}
The minimum of  $F(\tilde{\nu}_{ex},\tilde{\nu}_{in})$ should be
found in the domain specified by inequalities
$|\tilde{\nu}_{in}|\leq 1/2$ and $|\tilde{\nu}_{ex}|\leq 1/2$. The
function  $F(\tilde{\nu}_{ex},\tilde{\nu}_{in})$  has only one
extremum (or saddle point) at
$\tilde{\nu}_{ex}=\tilde{\nu}_{in}=0$. Therefore, its minimum is
reached either at the point of the extremum or at the edge of the
domain of allowed values of $\tilde{\nu}_{in}$,
$\tilde{\nu}_{ex}$. Thus, for determining the stability region of
the dimer phase it is sufficient  to find the minimum value of
expression (\ref{42}) at the edge of the domain and compare it to
the energy of the dimer phase in the absence of charge ordering,
$F_d=-(I+J)/4$.

We obtain that for $WN < 2I$ the minimum of (\ref{42}) at the edge
of the domain is reached for
$\tilde{\nu}_{in}=\tilde{\nu}_{ex}=\pm 1/2$. Since $F(\pm 1/2, \pm
1/2)=W/4-I/2 > F_d$ for such parameters the dimer state is stable
against the transition to the charge ordered state.

For $WN
> 2I$, the minimum at the edge domain is reached at the points
\begin{equation}\label{43}
\tilde{\nu}_{ex}=\pm\frac{1}{4}\frac{N-2}{N-1-\frac{I}{W}}, \qquad
\tilde{\nu}_{in}=\pm \frac{1}{2},
\end{equation}
where the value of the energy is equal to
\begin{equation}\label{44}
  \frac{I}{2N}-\frac{I}{2}
  +\frac{W(W N-2I)(N-2)}{2N( W N + I -W)}.
\end{equation}
In particular, it follows from (\ref{43}) that for $N\to\infty$
the minimum of energy at the edge of the domain corresponds just
to the state mentioned above
($\tilde{\nu}_{ex}=\tilde{\nu}_{in}/2=1/4$).

The value of expression (\ref{44}) can be larger or smaller than
$F_d$. Therefore, there is a region of instability of the dimer
phase in the space of parameters $(d,N)$. It is determined by the
system of inequalities
\begin{eqnarray}\label{45}
&&\frac{I}{2N}-\frac{I-J}{4}+\frac{W(W N-2I)(N-2)}{2N( W N + I
-W)}<0\cr&&WN>2I.
\end{eqnarray}

\begin{figure}
\begin{center}
\includegraphics[width=8cm]{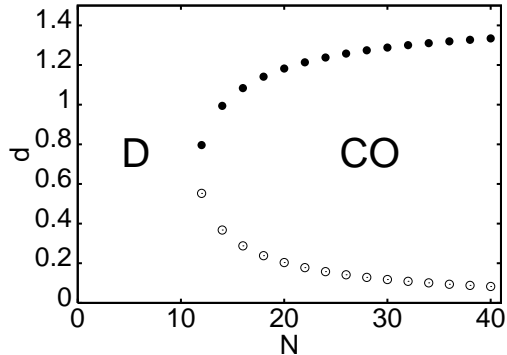}
 \caption{\label{fig1} Domains of existence of the dimer (D) and charge-ordered (CO) states and
 $d_{c1}$ ( $\circ$), $d_{c2}$ ($\bullet$) as functions of the number of layers $N$.
 The interlayer distance $d$ is given in units of the magnetic length $\ell_B$ . }
 \end{center}
\end{figure}

We analyze the system of inequalities  numerically and find that
for $N\leq10$ the conditions (\ref{45}) are not satisfied for any
$d$. Consequently, in a multilayer system with even $N\leq 10$ the
dimer state is stable against transition to the charge-ordered
phase at any values of $d$. For $N > 10$ the inequalities
(\ref{45}) are satisfied in certain interval of $d$: $d_{c1} < d <
d_{c2}$. At such $N$ and $d$ there will be complete charge
ordering of the inner layers and no phase coherence between those
layers. The dependence of $d_{c1}$ and $d_{c2}$ on $N$ is shown in
Fig. \ref{fig1}. Since $\lim_{N\to\infty} d_{c1}(N)=0$, for
$N\to\infty$ the domain of values $d$ for which the dimer state
can exist is restricted by the condition $d
> 1.45\ell_B$. However, at large but finite $N$
there exists a region of small $d$ ($d\lesssim
\ell_B\sqrt{2\pi}/N$) in which the formation of the dimer state is
possible.

To conclude this section we discuss shortly the role of tunneling.
It was shown in Ref. \onlinecite{25} that the interlayer tunneling
unfavours the charge ordering in multilayer systems. Based on a
comparison of the energy of the system for $\tilde{\nu}=0$ and
$\tilde{\nu}=\pm 1/2$ the authors of that paper arrived at the
conclusion that in the case when the tunneling parameter $t$
exceeds  $0.05 e^2/\varepsilon\ell_B$ the charge ordering becomes
energetically unfavorable for any values of the parameter
$d/\ell_B$ . In the presence of tunneling between layers the phase
coherence is no longer spontaneous, since tunneling leads to
fixing of the phase of the order parameter. In that case the
superfluid state with a constant phase gradient cannot be
realized, but superfluidity of a "soliton" type can exist
\cite{16,26,27}. But, according to the results of Ref.
\onlinecite{16}, already for $t\gtrsim 0.01 e^2/\varepsilon\ell_B$
the Landau criterium is not fulfilled at any superfluid
velocities. Therefore,  tunneling plays more of a negative than a
positive role for  realization of the superfluid state in
multilayer systems,. On the contrary, the use of multilayer
systems with parameters lying in the stability region of the dimer
state (Fig. \ref{fig1}) and with a small value of the tunneling is
more promising.

\section{CONCLUSION}

We have shown that a state with spontaneous interlayer phase
coherence can be realized in multilayer quantum Hall systems in a
certain range of parameters. The phase coherence in such systems
has a two-dimensional rather than a three-dimensional character;
namely, the system separates into pairs of adjacent layers, and
coherence is established only within each pair, and coherence is
absent between electrons belonging to diffierent pairs. If the
number of layers $N\leq 10$, then for an arbitrary distance
between layers the state with spontaneous interlayer phase
coherence is stable against transition to a charge-ordered phase.
If the number of layers $N>10$, the state with spontaneous phase
coherence is stable only for sufficiently small or sufficiently
large distances between layers. In the intermediate region of
interlayer distances in a system with more than 10 layers a
transition occurs to a charge-ordered state in which interlayer
phase coherence is absent.

Since the phase coherence has a two-dimensional character, a
multilayer quantum Hall system can be considered as a  solid-state
analog of ultracold atomic Bose gases in optical superlattices
(see, e.g., Ref. \onlinecite{28}). Among new effects that can
expect in such systems we would mention the nondissipative
interlayer drag of superfluid flow, in analogy with similar effect
predicted for bilayer Bose systems \cite{29,30,31}.

\section*{acknowledgment}
This study was supported by INTAS Grant No. 01-2344. Work was
begun at the Max Planck Institute of Physics of Complex Systems in
Dresden, and one of the authors (S.I.S.) takes this opportunity to
thank Prof. Fulde for the invitation to the Institute and for
hospitality.

\end{document}